\documentclass[
superscriptaddress,
 amsmath,amssymb,
 aps,
 prl,
 twocolumn,
floatfix,
longbibliography
]{revtex4-2}

\usepackage{amsmath}
\usepackage{amssymb}
\usepackage{amsfonts}

\usepackage[colorlinks,linkcolor=blue,anchorcolor=blue,citecolor=blue,urlcolor=blue]{hyperref}
\usepackage[utf8]{inputenc}
\usepackage{orcidlink}

\usepackage{xcolor}

\usepackage{dsfont}

\usepackage{braket}
\usepackage{graphicx}
\usepackage{dcolumn}
\usepackage{bm}

\renewcommand{\vec}[1]{\mbox{\boldmath $#1$}}

\begin{document}

\preprint{APS/123-QED}

\title{Description of the proton-decaying  0$^+_2$ resonance of the \boldmath{$\alpha$} particle}

\author{N. Michel\orcidlink{0000-0003-4214-8620}}
\affiliation{AS Key Laboratory of High Precision Nuclear Spectroscopy, Institute of Modern Physics, Chinese Academy of Sciences, Lanzhou 730000, China}
\affiliation{School of Nuclear Science and Technology, University of Chinese Academy of Sciences, Beijing 100049, China}
\affiliation{Grand Acc\'el\'erateur National d'Ions Lourds (GANIL), CEA/DSM - CNRS/IN2P3, BP 55027, F-14076 Caen Cedex, France} 

\author{W. Nazarewicz\orcidlink{0000-0002-8084-7425}}
\affiliation{Facility for Rare Isotope Beams and Department of Physics and Astronomy, Michigan State University, East Lansing, Michigan 48824, USA}

\author{M. P{\l}oszajczak\orcidlink{0000-0001-5206-0273}}
\affiliation{Grand Acc\'el\'erateur National d'Ions Lourds (GANIL), CEA/DSM - CNRS/IN2P3, BP 55027, F-14076 Caen Cedex, France}

\date{\today}

\begin{abstract} 
The recent precise experimental determination of the  monopole transition form factor from the ground state of $^4$He to its  $0^+_2$ 
resonance via electron scattering   has reinvigorated discussions about the nature of this first excited state of the $\alpha$ particle. The  $0^+_2$ state has been traditionally interpreted in the literature as
the isoscalar monopole resonance (breathing mode) or, alternatively, as a particle-hole shell-model excitation. To better understand the nature of this state, which lies only  $\sim$410\,keV above the proton emission threshold, we employ the coupled-channel representation of the no-core Gamow shell model. By considering the $[^3$H$ + p]$, $[^3$He$ + n]$, and $[^2$H$+^2${\rm H}$]$ reaction channels, we explain the excitation energy and monopole form-factor of the $0^+_2$ state. We argue that the continuum coupling  strongly impacts the nature of this state, which  carries characteristics of the proton decay threshold.

\end{abstract}

\maketitle

\textit{Introduction --} The lowest excited state of the $\alpha$ particle is the $0^+_2$ state at $E=20.1$\,MeV. Since 1966 \cite{Werntz1966} this state has been associated with
a  vibrational monopole  (breathing) mode 
resulting from the $0s\rightarrow 1s$ particle-hole 
shell-model excitations
\cite{Blomqvist1967,Szydlik1970,Flocard1975,Abgrall1975,Burov1981,Carvalho1987}.  
It was noted, however,  that two- and
four-particle excitations from $0s$ to $0p$ shell could also account for the low-lying positive parity excited
states, including $0^+_2$, hence the vibrational interpretation may not be strictly valid \cite{Schwesinger1981,Liu1984,Vassanji1988}.

The $0^+_2$ state lies 0.41\,MeV above the proton decay threshold and hence it decays 100\% by proton emission \cite{ensdf}. The $[^3$He$ + n]$  decay threshold lies only 0.37 keV above,
and the  $[^2$H$+^2${\rm H}$]$ threshold lies  3.64\,MeV higher \cite{Tilley1992}. Consequently, the  theoretical description of the $0^+_2$ state  must consider the coupling to the nearby decay thresholds, i.e., it requires an open-quantum-system (OQS) framework.

An early description of nucleon reaction channels in $^4$He was provided in Ref.~\cite{Halderson1979} in terms of the continuum shell model.  Excited states of $^4$He were studied in
Ref.\,\cite{Carlson1984} using variational R-matrix approach with realistic interactions.
The four-body bound-state calculations of $0^+_2$ interpreted this state as $^3$H+$p$ configuration \cite{Hiyama2004}, and provided a quantitative  description od  the  monopole
$0^+_1\rightarrow 0^+_2$ form factor (for supporting discussion,
see also Ref. \cite{Horiuchi2008}). 
In Refs.~\cite{Bacca2013,Bacca2015}, the Lorentz integral transform    approach, implicitly taking into account continuum effects, 
has been used to describe the $0^+_2$ state and the monopole form factor. The  results turned out to be  strongly dependent on the interaction used (see also discussion in Refs~\cite{Viviani2020,Kegel_2023}). 
The importance of the threshold structure on the properties of the $0^+_2$ state was  demonstrated  in Ref.~\cite{Kirscher2018} in the framework of pionless effective field theory.
Let us also mention significant literature on the microscopic description of four-nucleon
scattering with realistic interactions
using the Faddeev-Yakubovsky  \cite{Lazauskas2020}, hyperspherical harmonics \cite{Kievsky2008,Viviani2020}, and Alt-Grassberger-Sandhas  \cite{Deltuva2007a,Deltuva2007} frameworks, and recent {\it ab initio} no-core shell model (NCSM)  studies combined with the resonating group method (RGM) \cite{kravvaris2020ab}. All these approaches can take into account channel coupling effects in a fully microscopic way.

Recently, the A1 Collaboration at Mainz Microtron (MAMI) has carried out the precise measurement of the $0^+_1 \rightarrow 0^+_2$ transition via inelastic electron scattering \cite{Kegel_2023}. The theoretical analysis carried out in their study confirmed the previous conclusions \cite{Bacca2013,Bacca2015} that modern nuclear forces fail to reproduce the data. However,
as noted in the accompanying  viewpoint \cite{Epelbaum2023},
a possible explanation could involve  ``the relationship between the form factor and the location of the $0^+_2$ resonance.'' 

A microscopic description of nuclear states close to the particle decay threshold requires the unitary formulation of the problem that would guarantee the flux conservation.
As shown in Refs.~\cite{Okolowicz2012,Okolowicz2013}, in  the models that obey unitarity,  the collective mixing of shell model eigenstates around the threshold gives rise to the appearance
of the so-called {\it aligned state} that  shares many features of the decay channel. In fact, there exist
many near-threshold resonances that exhibit a large degree of resemblance to the nearby reaction threshold (for illustrative recent examples, see Refs.~\cite{Okolowicz2020,Okolowicz2023}).
Considering the above, it is indeed tempting to interpret the 
  $0^+_2$ excitation of the $\alpha$ particle as an eigenstate aligned with respect to the $[^3$H$ +p]$ threshold.
To this end, 
we employ the no-core Gamow shell model (GSM) \cite{PhysRevC.88.044318,Li2021} in the coupled-channel representation  (NCGSM-CC) \cite{michel_book_2021}. The  unitary OQS formalism of the NCGSM-CC  is very well suited for this task as it is based on realistic input and  is capable of describing reaction channel coupling effects.

\textit{Coupled-channel representation of the no-core Gamow shell model --} Here we will briefly outline the essential steps in the calculation of the monopole form-factor of the $0^+_2$ resonance using the no-core NCGSM-CC formalism. More details about the NCGSM-CC approach can be found in the Supplemental Material \cite{Sup}.   

In NCGSM-CC, the ${ A }$-body state  is decomposed into reaction channels defined as binary clusters  (target T and projectile P) involving different numbers of neutrons and protons.
The binary-cluster channel states are defined as:
\begin{equation}
\ket{ \left( c , r \right)_{ M }^{ J } } = \hat{ \mathcal{A}} [\ket{J^{(\rm int)}_{\rm T}} \otimes  \ket{r ~ \ell ~ J^{(\rm int)}_{\rm P} ~ j}]_{ M }^{J} \label{channel}
\end{equation} 
where the channel index $c$ stands for different quantum numbers and mass partitions, ${\hat{ \mathcal{A}}}$ is the inter-cluster antisymmetrizer that acts among the nucleons pertaining to different clusters with    angular momenta ${J_{ \rm T } }$ and ${J_{ \rm P } }$.
The total angular momentum is ${ \vec{J} = \vec{j} + \vec{J}^{(\rm int)}_{\rm T} } $ where $\vec{j} = \vec{\ell} + \vec{J}^{(\rm int)}_{\rm P}$.

All coordinates and angular quantum numbers in (\ref{channel}) are defined with respect to the center of mass (c.m.)~of the target+projectile composite. Since the target state $\ket{J^{(\rm int)}_{\rm T}}$ is intrinsic,  it has no c.m.~component. The projectile state $\ket{r ~ \ell ~ J^{(\rm int)}_{\rm P} ~ j}$ contains both the intrinsic wave function of the projectile $\ket{{ J }_{ \rm P }^{\rm (int)} }$ and the relative wave function $\ket{r ~ \ell}$, which can be identified with the projectile's c.m.~motion in the asymptotic zone.
By  expanding the radial wave function $u_c(r)$ in the single-particle (s.p.)  Berggren basis $u_n(r)$ \cite{BERGGREN1968265,michel_book_2021} that involves resonant states and scattering continuum,  one can derive the standard RGM expressions \cite{wildermuth_tang_book_1977,michel_book_2021}  for the Hamiltonian and norm kernels \cite{Sup}).

To handle the transformation from laboratory coordinates to c.m.~and relative coordinates, it is  convenient to use the harmonic oscillator (HO) basis. We emphasize that this procedure is consistent with the use of the Berggren expansion as the HO basis is used only to calculate the finite-range parts of NCGSM-CC potentials. The full NCGSM-CC coupled-channel Hamiltonian is diagonalized in the Berggren basis, so that the weakly bound and resonant eigenstates have proper asymptotic behavior.

In the first step,  we calculate the ground states of target and projectile. 
These wave functions can be represented as the products of the intrinsic wave functions $ \ket{J^{(\rm int)}}$  and the c.m.~HO wave functions $ \ket{0s_{\rm CM}}$  obtained by using the standard Lawson method \cite{Lawson}. 
To suppress the spurious c.m.~component in the composite product states built from  projectile and target states, an additional Talmi-Moshinsky-Brody \cite{Talmi1952,Moshinsky1959,Brody_book_1965} transformation has to be performed, see \cite{Sup}. 
As this channel state is a linear combination of Slater determinants, the  matrix elements of $\hat{ H }$ are straightforward to calculate. The numerical precision of the c.m. motion removal  has been checked  by explicitly acting on the channel state with the  c.m.~Hamiltonian; the resulting c.m. mode suppression is  $\sim$$10^{-11}$ or smaller.

The many-body matrix elements in NCGSM-CC are calculated using the Slater determinant expansion of the cluster wave functions. The RGM treatment of the non-orthogonality of channels is standard\,\cite{michel_book_2021}.
Note that the antisymmetry of channels, enforced by the antisymmetrizer in (\ref{channel}), is exactly taken into account through the Slater-determinant expansion of many-body target and projectile states.

\textit{Results --} In the following, we apply the NCGSM-CC framework to the structure of the proton-unstable resonance $0^+_2$ of $^4$He and the related monopole transition form factor $F_{rel}(q^2)$. 
To this end,  we  study  the channel  occupation $\mathcal{R}e(a_c^2)$   corresponding to the NCGSM-CC wave function  of the $0^+_2$ state, where $a_c^2$ is the squared norm of the channel wave function. The channel occupation and its variation as a function of the energy $\Delta{E}_{\rm th}
=E-E_{\rm thr}$ from the $[^3$H$ + p]$  threshold  provide insights into   the microscopic structure of the $0^+_2$ resonance.

The NCGSM Hamiltonian used is based on  the $V_{low-k}$ \cite{V_low-k} N$^3$LO interaction of Entem and Machleidt with  $\Lambda = 1.9$ fm$^{-1}$ \cite{PRC_N3LO}. A model space of $12 \hbar \omega$ is used for the 
%therein, which forms either the total space of NCSM, or the 
HO expansion of channel functions which is sufficiently large to provide a satisfactory description of both $0^+_1$ and $0^+_2$  states of $^4$He. 
In our analysis, we consider three  channels: $[^3$H$+p]$, $[^3$He$+n]$, and $[^2$H$+^2${\rm H}$]$, corresponding to three different mass partitions of $^4$He. The intrinsic wave functions of $^3$H, $^3$He and $^2$H clusters are assumed to correspond to their ground states. The relative partial waves included in the basis are $s_{1/2}$ ($[^3$H$+p]$, $[^3$He$+n]$), and $^3S_1$, $^3D_1$ ($[^2$H$+^2${\rm H}$]$). The Berggren basis contours in the complex momentum plane are defined by the linear momenta $k = 0$, $k_{\rm peak} = 0.2-0.19i$, $k_{\rm middle} = 0.4-0.19i$, and $k_{\rm max} = 3$ (all in fm$^{-1}$); they are discretized with 90 points, to which the poles $0s_{1/2}$,  $1s_{1/2}$ ($[^3$H$+p]$, $[^3$He$+n]$), the three and two first resonant states of the $^3S_1$ and $^3D_1$ partial waves ($[^2$H$+^2${\rm H}$]$) are added, respectively. This guarantees that the  resonance $0^+_2$ is properly described in the asymptotic zone. 
Since the $^3$H, $^3$He and $^2$H clusters are assumed to be in their  positive-parity ground states, the $\ell = 1$ partial waves in projectile wave functions are not allowed.

The three-body part of the N$^3$LO interaction has been neglected.  Based on our NCGSM calculations,  the effect of three-body forces  on the binding energy is about 1.5 to 2 MeV.  However, as discussed in Ref.~\cite{Carlson1984}, three body force  has a very small effect on the spectrum above the proton decay threshold. Moreover, it is always necessary to  modify the Hamiltonian by fixing the positions of particle-emission thresholds as it is impossible to reproduce binding energies up to a required precision of a few keV. In this way, the small energy shift due  three-body force is phenomenologically accounted for.
%, which is consistent with other approximations assumed in  our study.
In all calculations we use the HO length of $b_{\rm HO} = 1.8$\,fm, as this value is optimal for the $0^+_2$ state and close to optimum for the $0^+_1$ ground state. For this choice of parameters, the energy of the $0^+_1$ ground state is about -25 MeV with NCGSM-CC, which is sufficiently precise for our study.

\begin{figure}[!htb] 
\includegraphics[width=1.\linewidth]{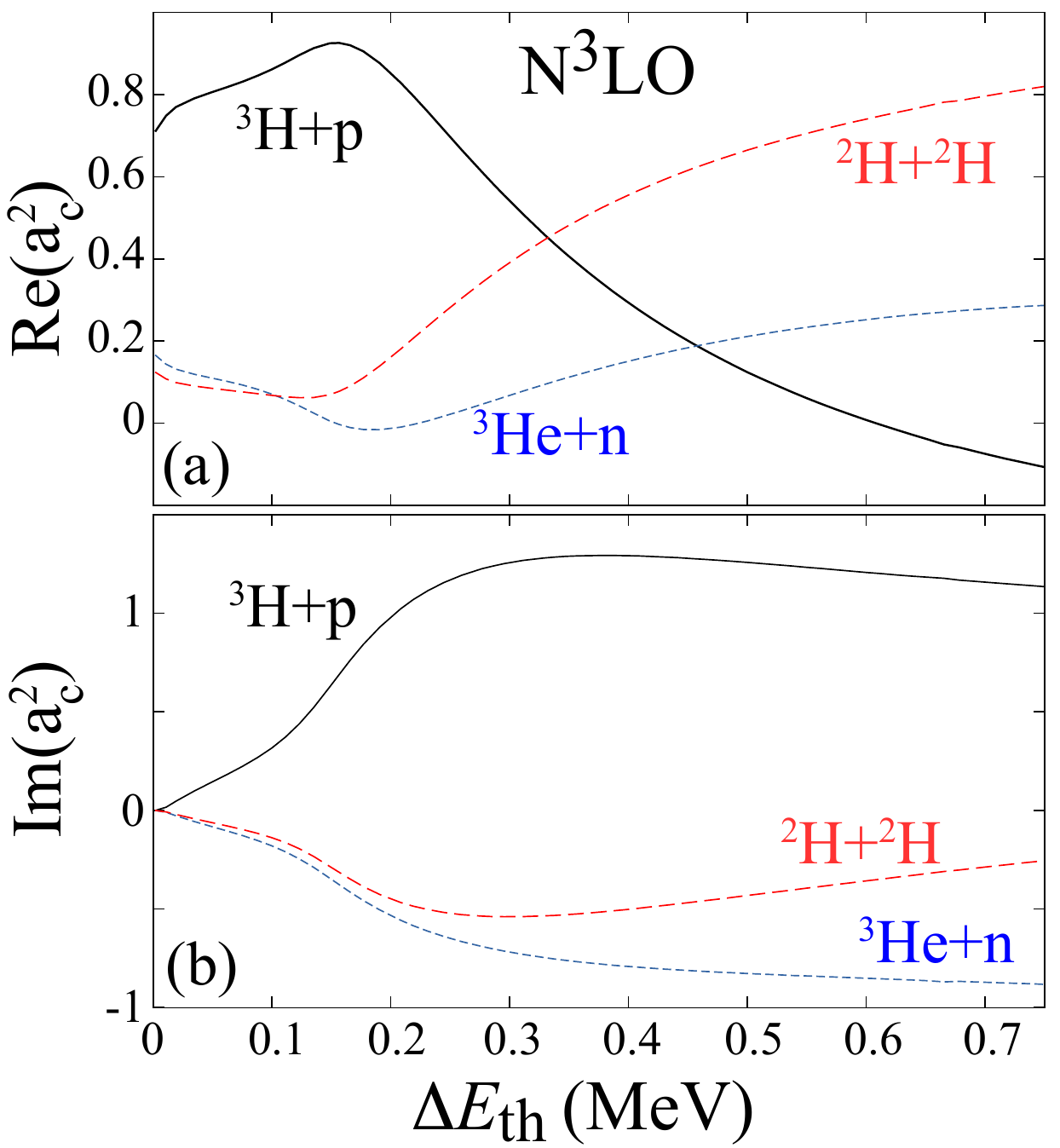} 
  \caption[T]{\label{Re_uc_occupation} Top:  occupations $\mathcal{R}e(a_c^2)$ of orthogonalized proton (solid line), neutron (dashed line), and deuteron channels (long-dashed line)  plotted as a function of the difference $\Delta E_{\rm thr}$ between the energy of the $0^+_2$ state and the proton threshold.
  Bottom: Corresponding imaginary channel occupations $\mathcal{I}m(a_c^2)$.}
\end{figure}

The  occupations $\mathcal{R}e(a_c^2)$  of proton, neutron, and deuteron channels are plotted in Fig.~\ref{Re_uc_occupation}(a). Note that RGM-orthogonalized channels are used in order to simplify the interpretation  \cite{michel_book_2021}.  This does not change asymptotic physical properties as the initial and RGM-orthogonalized channels are identical at large distances. 
Therefore,  the channel label used in Fig.~\ref{Re_uc_occupation} properly characterizes the asymptotic behavior.

As shown in Fig.~\ref{Re_uc_occupation}(a),  the [$^3$H$+p$] channel is dominant at $E <320$\,keV. For $\Delta E_{\rm thr}\simeq 150$\, keV, the percentage of the [$^3$H$+p$] channel 
is $\sim$95\%, whereas the weights of closed channels [$^3$He$+n$] and $[^2$H$+^2${\rm H}$]$ are very small. At the  experimental value of $\Delta E_{\rm thr}\approx 400$\,keV,  the [$^3$H$+p$] channel occupation drops to 32\% and the  $[^2$H$+^2${\rm H}$]$ channel  becomes dominant. The neutron decay channel opens up at $\Delta E_{\rm thr}\approx 0.78$\,MeV. At this energy, the proton-channel amplitude is very small while the wave function is dominated by the deuteron-deuteron component. 
It is to be noted that in the complex-energy framework such as GSM,  the squared wave-function amplitude  $\mathcal{R}e(a_c^2)$ can be negative \cite{michel_review_2009}. 
The statistical uncertainty of the $\mathcal{R}e(a_c^2)$  which, at the leading order, is associated with its imaginary part $\mathcal{I}m(a_c^2)$ \cite{Berggren1996,michel_book_2021} is shown in  Fig.~\ref{Re_uc_occupation}(b).
A statistical uncertainty on $\mathcal{R}e(a_c^2)$ arises because of the different life times that the resonance 0$^+_2$ state can have in several experiments. Hence, as explained in 
Refs.\,\cite{Berggren1996,michel_book_2021,Myo2023}, $\mathcal{R}e(a_c^2)$ is the average value of the corresponding occupation probability obtained in different measurements, while $\mathcal{I}m(a_c^2)$  can be related to the dispersion rate over time in the measurement, and hence represents its statistical uncertainty.
The imaginary squared amplitudes  are of the same order of magnitude as $\mathcal{R}e(a_c^2)$, except for a narrow region above the proton-emission threshold. Thus, there is a large statistical uncertainty on channel occupations in the energy region considered.
This is consistent with the fact that 
the measured spectral function of the $0^+_2$ state deviates from the Breit-Wigner profile
\cite{Kegel_2023}, i.e., 
this state cannot be considered as quasi-stationary \cite{wang2023,Myo2023}.
The NCGSM-CC predicts a large decay width,  $\Gamma \approx 1200$\,keV at $\Delta E_{\rm thr}\approx 400$\,keV. Consequently, the predicted $0^+_2$ state is very broad. This means that it becomes difficult to relate the value of $\Gamma$, associated with the imaginary part of the complex energy  of  the $0^+_2$ state with the lifetime, see   Ref.~\cite{wang2023}.

\begin{figure}[!htb] 
\includegraphics[width=1.\linewidth]{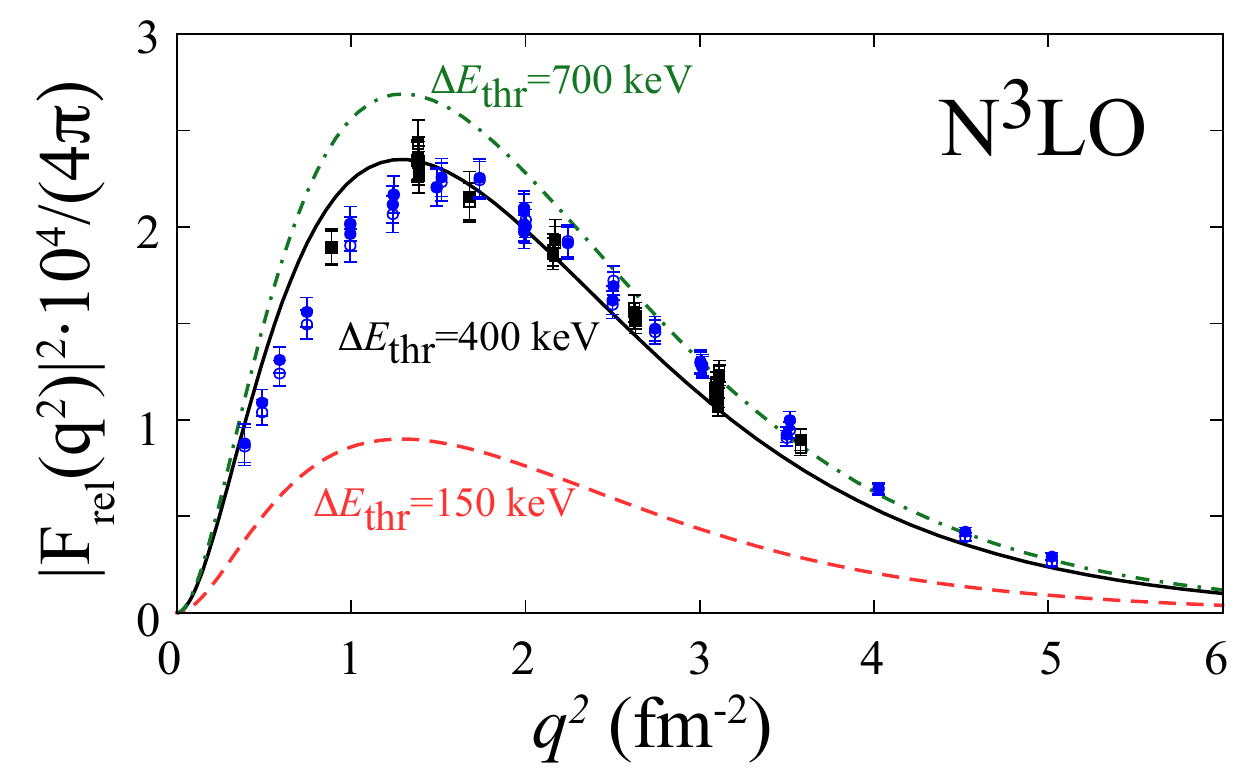}
  \caption[T]{\label{Frel_q2} Monopole transition form factor $|F_{\rm rel}(q^2)|^2$ obtained with N$^3$LO for the excitation of  the ground state $0^+_1$ to the  excited  $0^+_2$ state of $^4$He as a function of $q^2$ for three values of $\Delta E_{\rm thr}=150$\,keV, 400\,keV and 700\,keV. For the experimental data, see  Ref. \cite{Kegel_2023}.}
\end{figure}
We  now consider the monopole transition form factor $|F_{rel}(q^2)|^2$,
whose definition can be found in Refs.\cite{Hiyama2004,Sup}.
We include the effects of the intrinsic charge
proton form factor and recoil terms (see Refs.\cite{Hiyama2004,Sup} for details).
Neutron finite-size effects are neglected as neutron form factor and meson exchange terms are small \cite{GALSTER1971221,Hiyama2004}. The monopole form factor $|F_{rel}(q^2)|^2$ is shown in Fig. \ref{Frel_q2} at $\Delta E_{\rm thr}=150$\,keV, 400\,keV and 700\,keV. 
 The 400\,keV  value corresponds to the experimental position of the $0^+_2$ resonance. It is seen that (i) the calculated magnitude of $|F_{rel}(q^2)|^2$  strongly depends on the energy distance from the proton threshold, and (ii) the NCGSM-CC calculation at the experimental value of $\Delta E_{\rm thr}$ reproduces experiment fairly well. It should be noted that the occupation of the channels [$^3$H$+p$] and [$^2$H$+^2$H] is comparable at $\Delta E_{\rm thr}=400$\,keV, suggesting a complicated continuum structure of the $0^+_2$ resonance. Moreover, the dispersion of the [$^3$H$+p$] channel occupation is maximal at this energy. 

\begin{figure}[!htb] 
\includegraphics[width=1.\linewidth]{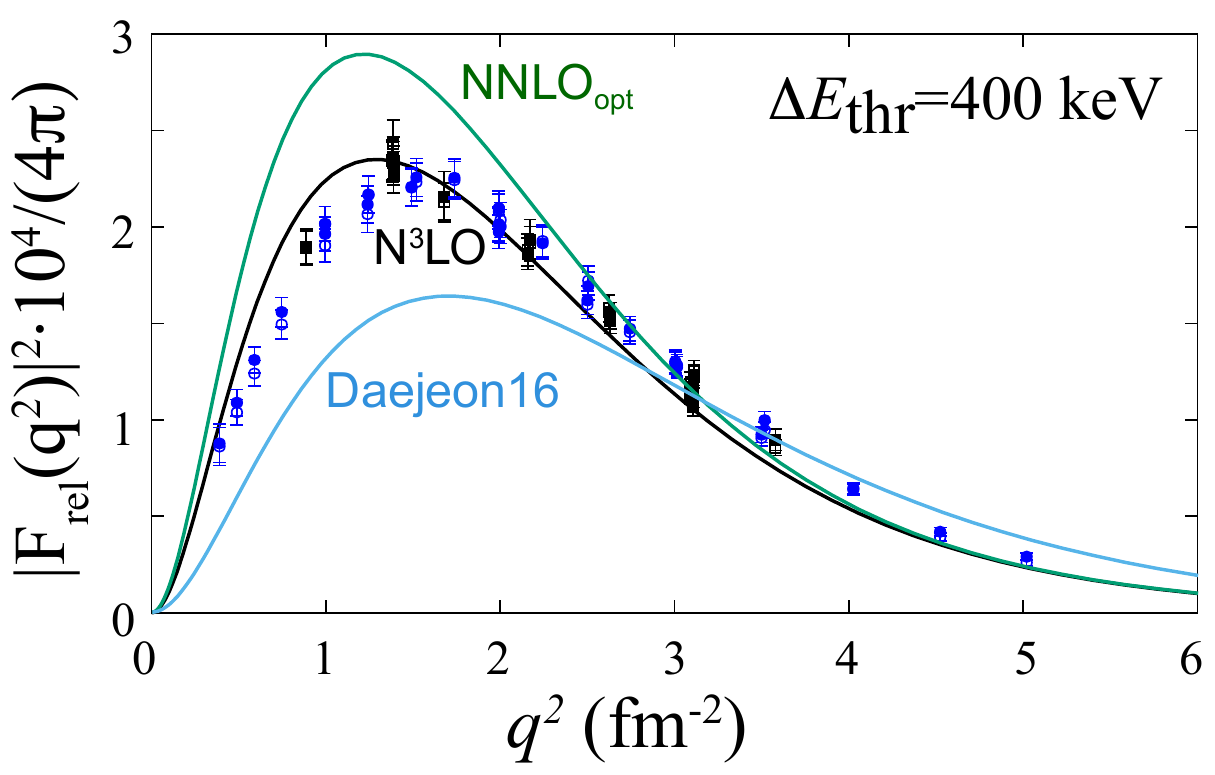}
  \caption[T]{\label{F_q2_three_interactions} Monopole transition form factor for the $0^+_1\rightarrow 0^+_2$ excitation  of $^4$He at $\Delta E_{\rm thr}=400$\,keV obtained with three  interactions: N$^3$LO \cite{V_low-k}, NNLO$_{\rm opt}$  \cite{PhysRevLett.110.192502}, and Daejeon16 \cite{PhysRevC.100.054329}.}
\end{figure}

In order to check the sensitivity of predictions to the choice of interaction, we carried out
calculations with NNLO$_{opt}$  \cite{PhysRevLett.110.192502} and Daejeon16 \cite{PhysRevC.100.054329} forces (see Fig.\ref{F_q2_three_interactions}).
The  form factors obtained with NNLO$_{\rm opt}$ and N$^3$LO interactions are fairly similar, with  the NNLO$_{\rm opt}$  prediction slighly overshooting the data at $q^2\approx 1.2$\,fm$^2$.
The  form factor obtained with Daejeon16 is lower in this region of $q^2$.
\begin{figure}[!htb] 
\includegraphics[width=1.\linewidth]{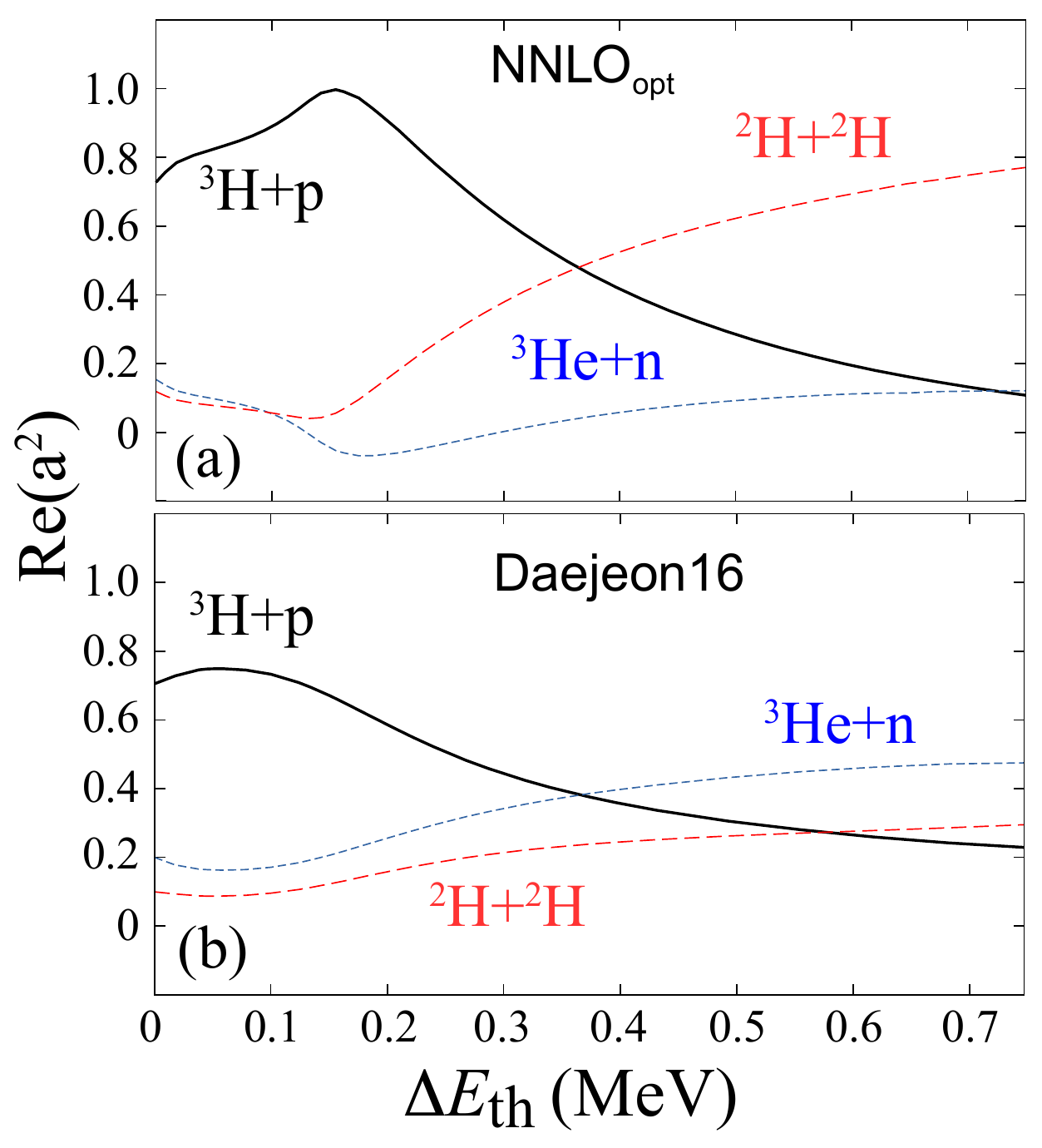}  
  \caption[T]{\label{Re_uc_occupation1} 
  Similar as Fig.\ref{Re_uc_occupation}(a) but for  (a) NNLO$_{\rm opt}$   and (b) Daejeon16. }
\end{figure}

The   NNLO$_{\rm opt}$  and Daejeon16  occupations $\mathcal{R}e(a_c^2)$ of orthogonalized channels are shown in Fig.~\ref{Re_uc_occupation1}.  One can see that the  difference of the occupations of 
the channels [$^3$H$+p$] and [$^3$He$+n$] at $\Delta E_{\rm thr}=400$\,keV 
 follows the increase of the monopole form-factor.

\textit{Conclusions --} In this Letter, stimulated by the recent experimental  study \cite{Kegel_2023}, we developed the OQS NCGSM-CC framework to understand the structure of 
the proton-decaying  0$^+_2$ excited  of the  $\alpha$ particle. Particular attention has been paid to the proper removal of the spurious c.m.~modes when constructing the channel wave functions. Our model based on the N$^3$LO chiral interaction reproduces the
binding energies of $0^+_1$ and   $0^+_2$ states,  a broad resonance  character of  the $0^+_2$ state, and the measured monopole form factor.

We predict a rather complex character of the $0^+_2$ state that involves a strong continuum coupling between the $[^3$H$ + p]$, $[^3$He$ + n]$, and $[^2$H$+^2${\rm H}$]$ decay channels.
The best agreement with the measured form factor  is obtained 
at the experimental energy $\Delta E_{\rm thr}$. We also predict a rather strong dependence of $|F_{rel}(q^2)|^2$ on the energy distance from the proton threshold, i.e., we answer the question raised in Ref.\cite{Epelbaum2023} affirmatively. In this respect, the 
 $0^+_2$ state should not be viewed in terms of a  breathing oscillation or
the $0s\rightarrow 1s$ particle-hole  excitation, but rather interpreted as  a threshold-aligned broad resonance whose structure is dominated by the interplay of the open channel $[^3$\rm H$ + p]$ and closed channels $[^3$\rm He$ + n]$ and $[^2$H$+^2${\rm H}$]$. Based on our  N$^3$LO results,
the explicit reproduction of competing particle-emission thresholds is important for the theoretical understanding of the  $0^+_2$ state. 

But the position of the $0^+_2$ state relative to the thresholds  is not the full story.
Calculations performed with other interactions, namely 
NNLO$_{\rm opt}$   and Daejeon16, have shown  some  interaction dependence of the 
wave function decomposition and hence  the
monopole form factor.
This suggests that -- when it comes to theoretical description --  the monopole form factor is a fairly sensitive and demanding observable: it is controlled by several factors, such as   interaction,  threshold positions, and  resonance energy.  Consequently, this quantity is not ideal when it comes to  constraining nuclear interaction.

\textit{Acknowledgments --}
We thank Sonia Bacca for providing experimental data and for useful discussion. N. Michel wishes to thank GANIL for the hospitality where this work has been done.
This material is based upon work supported by
the National Natural Science Foundation of China under Grant No.12175281, and the State Key Laboratory of Nuclear Physics and Technology, Peking University under Grant No. NPT2020KFY13, and by
the U.S.
Department of Energy, Office of Science, Office of Nuclear
Physics under Awards No. DOE-DE-SC0013365 (Michigan State University),  DE-SC0023175 (NUCLEI SciDAC-5 collaboration).

\bibliography{refs}

\end{document}